\documentclass[journal=jctcce,manuscript=article,layout=traditional]{achemso}

\pdftrailerid{}
\usepackage[T1]{fontenc} 
\usepackage{amsmath}
\usepackage{subcaption}
\usepackage{mathtools}
\usepackage{graphicx}
\usepackage{xcolor, soul}

\SectionNumbersOn

\title{A Simple Molecular Model for Hydrated Silicate Ionic Liquids, a Realistic Zeolite Precursor}

\author{Jelle Vekeman}
\affiliation[Ghent University]{Center for Molecular Modeling (CMM), Ghent University, Technologiepark-Zwijnaarde 46, 9052, Ghent, Belgium}
\author{Dries Vandenabeele}
\affiliation{Center for Surface Chemistry and Catalysis Characterisation and Application Team (COK-KAT), KU Leuven, Celestijnenlaan 200F, 3001, Leuven, Belgium}
\author{Nikolaus Doppelhammer}
\affiliation{Center for Surface Chemistry and Catalysis Characterisation and Application Team (COK-KAT), KU Leuven, Celestijnenlaan 200F, 3001, Leuven, Belgium}
\author{Elisabeth Vandeurzen}
\affiliation{Center for Surface Chemistry and Catalysis Characterisation and Application Team (COK-KAT), KU Leuven, Celestijnenlaan 200F, 3001, Leuven, Belgium}
\author{Eric Breynaert}
\affiliation{Center for Surface Chemistry and Catalysis Characterisation and Application Team (COK-KAT), KU Leuven, Celestijnenlaan 200F, 3001, Leuven, Belgium}
\alsoaffiliation{NMR-Xray Platform for Convergence Research (NMRCoRe), KU Leuven, 3001 Leuven, Belgium}
\author{Christine E.A. Kirschhock}
\affiliation{Center for Surface Chemistry and Catalysis Characterisation and Application Team (COK-KAT), KU Leuven, Celestijnenlaan 200F, 3001, Leuven, Belgium}
\author{Toon Verstraelen}
\affiliation[Ghent University]{Center for Molecular Modeling (CMM), Ghent University, Technologiepark-Zwijnaarde 46, 9052, Ghent, Belgium}
\email{toon.verstraelen@ugent.be}


\begin{document}

    \newpage
    \begin{abstract}
        Despite the widespread use of zeolites in chemical industry, their formation process is not fully understood due to the complex and heterogeneous structure of traditional synthesis media.
        Hydrated silicate ionic liquids (HSILs) have been proposed as an alternative.
        They are truly homogeneous and transparent mixtures with low viscosity, facilitating experimental characterization.
        Interestingly, their homogeneous nature and simple speciation brings realistic molecular models of a zeolite growth liquid within reach for the first time.
        In this work, a simple molecular model is developed that gives insight into the crucial role of the alkali cations (sodium, potassium, rubidium and cesium).
        Thereby molecular dynamics simulations are combined with experimental measurements to demonstrate that the HSIL liquid structure strongly depends on the charge density and concentration of the alkali cation.
        As the water content increases, it transitions from a glassy network with fast ion exchange to an aqueous solution containing long-lasting, solvated ion pairs.
        Furthermore, simulations reveal that the cation is capable of bringing several silicate monomers together in the glassy network, displaying perfect orientations for condensation reactions that underlie zeolite formation.
        This work is an important step towards the development of molecular models that can fully describe the early nucleation process of zeolites in combination with experiments.
    \end{abstract}

    \begin{tocentry}
        \includegraphics{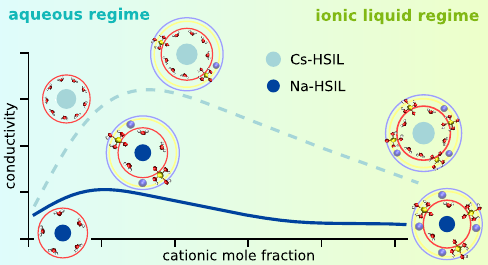}
    \end{tocentry}

    \newpage

    \section{Introduction}
    \label{sec:introduction}

    Zeolites are crystalline, nanoporous aluminosilicates representing the current workhorse of the chemical industry in the form of catalysts, molecular sieves and ion-exchangers.\cite{Li2021}
    Numerous key industries as well as fields with emerging significance rely on them, such as refinery and upgrading of fossil and bio-fuels,\cite{Perez-Botella2022} detergens,\cite{Koohsaryan2020} refrigeration and separation\cite{Feng2021}, water purification\cite{Wang2020}, desalination\cite{Algieri2021}, waste management\cite{Soudejani2019} and batteries.\cite{Chi2021}
    Nevertheless, the formation process of zeolites is still poorly understood due to the complex, heterogeneous sol-gel chemistry and hydrothermal conditions in traditional synthesis media.\cite{Rimer2018}
    Experimentally, most studies are restricted to a single, specific zeolite and require a costly combination of multiple state-of-the art techniques.\cite{Schwalbe-Koda2021}
    From a computational perspective, development of molecular models for such traditional growth liquids is problematic as \textit{ab initio} methods are too expensive to perform at the scale (both size and time) that is needed for nucleation, while classical methods struggle with appropriate representation of the complex, rapidly changing speciation.\cite{Carvalho2022}
    Due to the complexity of traditional zeolite growth liquids, prior studies have mostly relied on modeling isolated aspects of the synthesis liquid.\cite{VanSpeybroeck2015}
    Many studies have focused on modeling isolated oligomerization steps using DFT calculations in implicit solvent to deduce reaction networks, sometimes taking into account the influence of organic templates or cations.\cite{Freeman2020, Ciantar2021a, Prasad2022}
    Others have made valuable attempts to include explicit solvent to obtain a more realistic description of zeolite nucleation, yet these models cannot capture the complexity of sol-gel systems.\cite{Ho2023, Verstraelen2009, Shere2022}
    Finally, attempts have been made to describe crystal nucleation and growth at a larger scale by sacrificing atomic resolution.\cite{Carvalho2022, Hill2021}

    Very recently, a novel synthesis path was described through entirely transparent and homogeneous hydrated silicate ionic liquids (HSILs).\cite{vanTendeloo2015, Castro2014}
    These room temperature, molten salts are formed through coacervation of tetraethylorthosilicate by MOH (M = Na$^+$, K$^+$, Rb$^+$ or Cs$^+$) and have been used for the synthesis of more than 15 different zeolite topologies at low temperatures (<100°C).
    Interestingly, HSILs are very suitable for \textit{in situ} experimental analyses, such as (moving electrode) electrochemical impedance spectroscopy,\cite{Doppelhammer2020, Brabants2017} atomic force microscopy\cite{Houlleberghs2019}, nuclear magnetic resonance spectroscopy\cite{Pellens2022} and X-ray diffraction.\cite{Asselman2022}
    From these studies, it was found that HSILs display limited speciation with oligomers not much larger than tetramers, while being very stable and not exhibiting zeolite formation until addition of an aluminum source.
    Zeolite is formed with an efficiency of 100\% in terms of aluminum solubility, allowing for careful control and quenching of the formation process at any stage.\cite{Houlleberghs2019}
    Furthermore, the physicochemical state of the mixture is practically unaltered for small yields.
    The final topology of the zeolite is clearly linked to precursor composition and more specifically, the type and concentration of the alkali cation.
    Therefore, it is important to understand how all properties of the growth liquid are affected by cation concentration and charge density, as these properties may play a crucial role in the initial stages of zeolite nucleation, both in HSILs and traditional growth liquids.
    In fact, it is hypothesized that the HSIL conditions are very similar to the liquid phase encountered in traditional sol-gel synthesis media.\cite{Asselman2023}
    In addition to facilitating experimental characterization, the simplicity of the experimental system discussed in this work offers a unique opportunity to develop a theoretical model that closely matches experiment.

    Experimental conductivity measurements of HSILs as a function of cation concentration, revealed two distinct regimes: an aqueous phase at high water content and an ionic liquid phase at low water content.\cite{Pellens2022, Doppelhammer2021}
    In the former regime, the conductivity rises with increasing ion concentration as the amount of charge carriers increases, while in the latter the conductivity decreases with increasing ionic content.
    It was hypothesized that the observed conductivity decrease is caused by the formation of strong ion pairing as the low water content can no longer solvate the ionic species, giving rise to a dynamic network of alkali cations and silicate anions with ionic liquid properties.
    This is especially important since it was further shown that ion pairing plays a crucial role in the initial stages of zeolite formation by giving rise to the formation of so-called prenucleation clusters.\cite{Pellens2022}

    Furthermore, HSILs are interesting study cases in themselves, because they exhibit different physical properties (e.g. conductivity, viscosity, ...) only by changing the alkali cation (and its concentration).\cite{Asselman2022a}
    For fully aqueous solutions of alkali cations, it is well known that the local solvent structure surrounding the cations correlates with decreasing charge density and increasing radius along the Li$^+$, Na$^+$, K$^+$, Rb$^+$ and Cs$^+$ series. \cite{Mahler2012}
    Smaller cations (Li$^+$) strengthen the local hydrogen bonding network, while larger cations (K$^+$, Rb$^+$ and Cs$^+$) have the opposite effect, and Na$^+$ is considered borderline.
    HSILs can be used to investigate in detail whether these known trends can be safely extrapolated to the ionic liquid regime at high ionic concentration.
    Note that this is very relevant in the context of electrolytes for novel batteries where a detailed understanding of ion-ion and solvent-ion interactions, liquid structure and conductivity are highly important.\cite{Yao2023}
    Recently, the concentration dependence of the conductivity of LiPF$_6$ and LiPF$_4$ was suggested to arise from an interplay between ion dissociation and viscosity, similarly to the HSIL systems.\cite{Koo2023}
    The similar trends suggest that the current work and the forthcoming conclusions are transferable to highly concentrated electrolytes relevant to novel battery technology.

    In this work, a simple molecular model -- employing silicate monomers, MOH and water -- is proposed for extensive molecular dynamics (MD) simulations to characterize the liquid structure and conductivity of HSILs.
    Simulations were performed for four different systems, i.e. Na-HSIL, K-HSIL, Rb-HSIL and Cs-HSIL, at 9 different compositions to assess the influence of the cation type and concentration on the liquid structure and conductivity.
    Note that Rb-HSIL is underexplored experimentally due to its high cost, making it an interesting addition in this computational study.
    From these simulations, a detailed understanding of the liquid structure is obtained as a function of cationic charge density and concentration through calculation of conductivity, diffusion coefficients, coordination numbers, radial distributions and the extent to which the movement of cations and anions is correlated.
    The theoretical results are validated by experimental measurements on HSILs having the same compositions.

    \section{Methods}
    \label{sec:methods}

    \subsection{Theoretical Simulations}

    Classical MD simulations were performed in the NPT ensemble using a Langevin thermostat with a friction constant of 1 ps$^{-1}$ and a Monte Carlo barostat as implemented in the OpenMM package version 7.5.1.\cite{Eastman2017}
    Simulations were performed at 300 K and 1 bar over 100 ns, 20 ns of which for equilibration, using a time step of 1 fs.
    Snapshots were saved every 1 ps for post-processing using MDAnalysis and MDTraj.\cite{Gowers2016, Michaud-Agrawal2011, McGibbon2015}

    Following the conventions of the zeolite community, the simulated systems can be represented using the notation H$_2$O:SiO$_2$:MOH $n$:1:1, whereby $n$ = 260.5, 90, 31, 18.5, 11, 6.6, 4, 2.5 or 2.2.
    This coincides with cationic (and anionic) molar fractions of 0.004, 0.011, 0.031, 0.050, 0.080, 0.123, 0.180, 0.248 and 0.267, respectively.
    The systems were set up to contain at least 10 000 species to minimize size effects, using Packmol.\cite{Martinez2009}
    Only silicate monomers were considered and assumed to be singly deprotonated.
    Thereby, the proton removed from the silicate monomer is assumed to form a water molecule with a free OH$^-$ ion coming from MOH.
    Given the MOH/SiO$_2$ ratio of 1, this means that in practice no OH$^-$ ions are present in the system, while an extra water molecule is added for every silicate monomer.~\cite{Pellens2022}
    Simulations using other MOH/SiO$_2$ ratios were attempted, but were unreliable due to unphysical ion aggregation at low concentrations.
    More specifically, it was found that the presence of OH$^{-}$ anions caused the formation of MOH crystals at concentrations well below the solubility limit of the respective systems.
    Blazquez et al.~\cite{Blazquez2023} showed that such problems are due to the lack of complexity of simple force fields when modeling strong electrolytes in water.
    Development of a solution to this problem is well beyond the scope of the current work, so that only $n$:1:1 H$_2$O:SiO$_2$:NaOH compositions are considered.
    Furthermore, no aluminates were taken into account as aluminates initiate nucleation, which cannot be described by the force fields used in this work.
    This is justified by the experimental observation that HSILs are stable prior to aluminate addition.\cite{vanTendeloo2015}
    The exact compositions and periodic box sizes of all systems can be found in the supporting information.

    Several reactive \cite{Moqadam2015,Carvalho2022,Mahadevan2020,Nayir2019,Fogarty2010} and non-reactive \cite{Pedone2006,VanBeest1990,Hu2012,Sanders1984,Lopes2006} force fields exist for silicates, yet for this work a cheap (and therefore non-reactive) atomistic force field was needed that distinguishes protonated and deprotonated oxygens bound to silicon, while also properly describing interactions with free alkali cations.
    The force field developed by Emami et al.,\cite{Emami2014} provides such opportunity as it was developed for a silica surface in contact with sodium cations.
    Although this force field contains specific parameters for sodium, it is compatible with the sodium parameters from CHARMM with limited loss of accuracy according to the original publication.
    The latter feature is important as it provides parameters for K$^+$, Rb$^+$ and Cs$^+$, which are not present in the Emami force field.
    As such, the parameters for the Si(OH)$_3$O$^-$ species come from the work by Emami et al., while the parameters for H$_2$O (TIP3P)\cite{Jorgensen1983} and alkali cation come from CHARMM.\cite{Beglov1994, Won2012}
    This approach was validated through a series of simulations on Na-HSIL using exclusively Emami parameters, leading to good qualitative agreement with the results reported below.
    The intra- and intermolecular interactions were described by a simple CHARMM-like force field, including Lennard--Jones potentials for the dispersion interactions, Coulombic terms for electrostatic interactions and harmonic approximations for intermolecular bonding and angle terms.
    Lennard--Jones parameters for unlike atom types are determined using the Lorentz--Berthelot mixing rules.
    A cutoff of 12 \AA\ was used for both the van der Waals and the Coulombic term.
    All used parameters can be found in the in the forcefield file that is provided in the supporting information.

    Diffusion coefficients for atom $i$, $D_i$, were calculated using
    \begin{align}
        D_i &= \lim_{t \to \infty} \frac{1}{6t} \langle | \textbf{R}_i(t) - \textbf{R}_i(0) |^2 \rangle
    \end{align}
    where $t$ is time, $\textbf{R}_i(t)$ is the position of the $i$-th ion at time $t$ and the angle brackets denote the ensemble average.

    The conductivity, $\sigma$, was calculated using\cite{Kubisiak2020}
    \begin{align}
        \sigma &= \lim_{t \to \infty} \frac{e^2}{6tVk_{\mathrm{B}}T} \sum_{i,j} z_i z_j \langle \left[  \textbf{R}_{i}(t) - \textbf{R}_i(0) \right] \left[  \textbf{R}_j(t) - \textbf{R}_j(0) \right] \rangle
    \end{align}
    where $V$ is the volume of the simulation box, $k_{\mathrm{B}}$ is the Boltzmann constant, $T$ is temperature, $e$ is the elementary charge and $z_i$ is the charge of ion $i$.
    In practice, the conductivity was calculated by splitting up the trajectory in 50 subtrajectories to increase statistics.\cite{Kubisiak2020}
    Besides the direct calculation through this Einstein relation, conductivity was also estimated from the cationic and anionic diffusion coefficients using the Nernst--Einstein approximation\cite{Kubisiak2020}
    \begin{align}
        \sigma_\text{NE} &= \frac{e^2}{6tVk_{\mathrm{B}}T} \left( D_{\mathrm{M^+}} + D_{\mathrm{Si(OH)_3O^-}} \right)
    \end{align}
    with $M$ being Na, K, Rb or Cs.
    In all post-processing, the Si atom was used as a reference for the silicate anion.

    Radial distribution functions, $g_{ab}(r)$, between atoms $a$ and $b$ were calculated according to
    \begin{align}
        g_{ab}(r) = \left( N_a N_b \right)^{-1} \sum_{i=1}^{N_a} \sum_{j=1}^{N_b} \langle \delta | \textbf{r}_i - \textbf{r}_j| - r \rangle
    \end{align}
    where $N_a$ represents the number of atoms $a$ and $\delta$ is the Kronecker delta.
    Water was represented by its oxygen atom.

    \subsection{Experimental Measurements}

    \begin{table}
        \begin{center}
        \begin{tabular}{l|c c c}
            X$_\text{MOH}$ & Na-HSIL & K-HSIL & Cs-HSIL\\
            \hline
            0.004 & 247.49 & 268.35 & 259.04 \\
            0.011 & 91.35 & 90.07 & 90.24 \\
            0.030 & 30.88 & 30.99 & 30.99 \\
            0.049 & 18.57 & 18.43 & 18.49 \\
            0.077 & 11.02 & 11.02 & 10.93 \\
            0.116 & 6.58 & 6.59 & 6.62 \\
            0.167 & 4.00 & 4.02 & 4.01 \\
            0.222 & 2.51 & - & 2.53 \\
        \end{tabular}
        \end{center}
        \caption{Number of water molecules, $n$, present in the experimental samples with molar compositions of H$_2$O:SiO$_2$:MOH $n$:1:1.
        Native K-HSIL has a molar composition of H$_2$O:SiO$_2$:MOH 3.6:1:1, such that a cationic mole fraction of 0.222 cannot be reached.}
        \label{tab:compositions}
    \end{table}

    Experimental conductivity measurements were performed using the moving electrode electrochemical impedance spectroscopy (MEEIS) meter at a temperature of 298.15K. \cite{Doppelhammer2020, Doppelhammer2021}
    The measurements included an 8-sample dilution series for each of the 3 systems: Na-HSIL, K-HSIL and Cs-HSIL, while Rb-HSIL was excluded due to its high cost.
    The samples had a molar composition of H$_2$O:SiO$_2$:MOH $n$:1:1, whereby the respective values for $n$ are given in Table~\ref{tab:compositions}, and equilibrium of the speciation was achieved within minutes in all cases.
    Please note that the experimentally considered mole fractions are slightly different from the simulated ones due to the respective restrictions on each of the used methods.
    However, all systems have the same SiOH/MOH ratio, allowing the calculated/measured properties to be compared as a function of the cationic mole fraction.

    To prepare the 3 pure HSIL systems, tetraethyl orthosilicate (TEOS), alkali hydroxide (MOH) and deionized water were combined as described previously\cite{Asselman2022a}.
    Following the hydrolysis of TEOS, a phase separation occurred.
    The upper phase containing water and ethanol was discarded, while the lower phase consisting of HSIL was retained.
    The pure HSILs have a molar composition of H$_2$O:SiO$_2$:MOH $n$:1:1, whereby $n$ = 2.2, 3.6 and 2.2 for the Na-HSIL, K-HSIL and Cs-HSIL, respectively.
    Experimental molar masses were obtained through gravimetric analysis, whereby complete hydrolysis of TEOS was assumed with all EtOH in the upper phase and all cations and silica in the lower phase.
    Under that assumption, the only unknown is the water fraction, which can be determined by the weighing both phases as described in more detail elsewhere.\cite{Haouas2014, vanTendeloo2015, Pellens2022a}
    Subsequently, the pure HSILs were diluted to achieve the desired molar compositions.

    Conductivity measurements were performed using the method of moving electrode electrochemical impedance spectroscopy with a custom cell design as discussed in prior work.\cite{Doppelhammer2020}
    10 ml of sample liquid were loaded into the preheated cell.
    A time of 10 minutes was given for temperature equilibration before measurement.
    Impedance spectra were measured at 11 equally spaced electrode distances ranging from 4 to 8 cm, each 31 logarithmically spaced frequencies ranging from 1 kHz to 1 MHz.
    Measurements were performed in potentiostatic model with a voltage amplitude of 100mV.

    \section{Results and Discussion}

    \subsection{Aqueous versus Ionic Liquid Regime}

    \begin{figure}
        \begin{center}
            \includegraphics{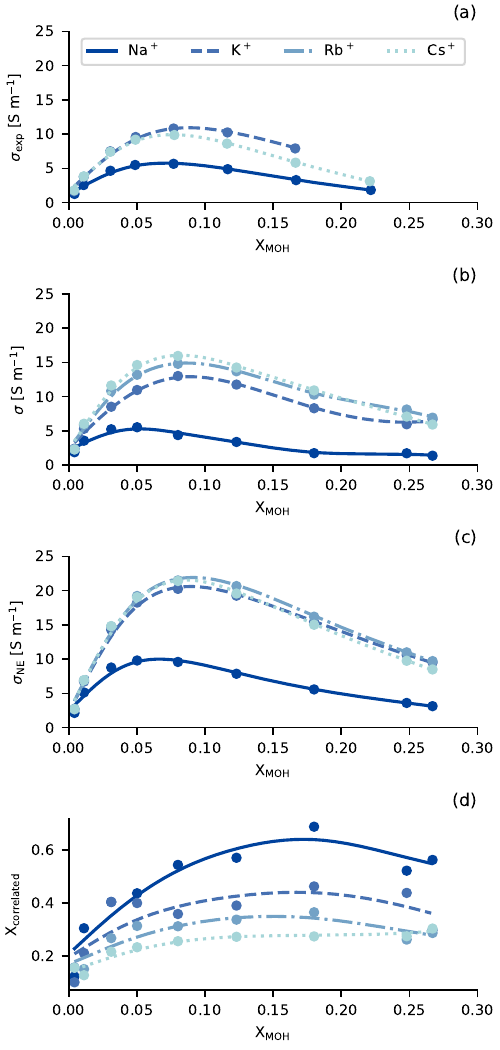}
        \end{center}
        \caption{(a) Experimental conductivity, (b) Theoretical conductivity, (c) Nernst--Einstein approximation and (d) fraction of correlated motion as a function of cationic mole fraction for Na-HSIL, K-HSIL, Rb-HSIL and Cs-HSIL.
        Error bars are at least an order of magnitude smaller than data points and thus omitted. The connecting lines are obtained through a spline fit to the data points.}
        \label{fig:conductivity}
    \end{figure}

    In prior work, the aqueous and ionic liquid regimes of HSILs were experimentally distinguished through measurement of the conductivity as a function of cation molar fraction.\cite{Pellens2022}
    The aqueous regime is characterized by increasing conductivity with increasing cation concentration, while the ionic liquid regime exhibits decreasing conductivity with increasing cation concentration.
    In this work, experimental conductivity measurements were performed for HSILs without aluminum for direct comparison with the theoretical predictions presented below.
    As shown in Figure~\ref{fig:conductivity} (a), the aqueous and ionic liquid regimes can clearly be distinguished from the experimental conductivity as a function of cationic molar fraction for all considered HSIL systems.
    A smoothing cubic spline is fitted to the data points to estimate the maximum values, using the make\_smoothing\_spline function from SciPy.\cite{Craven1978, 2020SciPy-NMeth}
    The splines reveal a maximum conductivity at a cationic mole fraction of 0.071 for Na-HSIL, 0.088 for K-HSIL and 0.076 for Cs-HSIL, while the maximum conductivities are 5.760 S m$^{-1}$, 10.932 S m$^{-1}$ and 9.924 S m$^{-1}$, respectively.
    Rb-HSIL is left out due to its high cost.
    These results are qualitatively similar to previously published results on HSILs containing traces of aluminum and a different water to cation ratio, whereby a maximum conductivity of 14.5 S m$^{-1}$  was observed at a cationic mole fraction of around 0.08 for a $n$:0.5:1:0.028 H$_2$O:SiO$_2$:NaOH:Al(OH)$_3$ HSIL system and of 23.63 S m$^{-1}$  around $X_{\text{Cs}}$ = 0.10 for a $n$:0.5:1:0.083 H$_2$O:SiO$_2$:CsOH:Al(OH)$_3$ HSIL system at 298.15 K.\cite{Pellens2022, Doppelhammer2021}

    In Figure~\ref{fig:conductivity} (b), theoretical results for the conductivity are given for the Na-HSIL, K-HSIL, Rb-HSIL and Cs-HSIL in the studied $n$:1:1 H$_2$O:SiO$_2$:NaOH compositions.
    The calculated conductivities display a maximum in all cases at similar cationic mole fractions as in experiment, while the conductivity for K-HSIL and Cs-HSIL is also substantially higher than for Na-HSIL.
    More specifically, a maximum conductivity for Na-HSIL is found of 5.29 S m$^{-1}$  at a cationic mole fraction of 0.052, while a maximum of 16.04 S m$^{-1}$  was found for Cs-HSIL at a cationic mole fraction of 0.082.
    The values for K-HSIL and Rb-HSIL are intermediate with conductivities of 12.91 S m$^{-1}$ at a cationic mole fraction of 0.089 and 14.91 S m$^{-1}$ at a cationic mole fraction of 0.086, respectively.
    Overall, it is clearly visible that the conductivity increases with increasing cationic radius, i.e. decreasing charge density.
    Furthermore, a large gap is seen between Na-HSIL conductivities and K-HSIL, Rb-HSIL and Cs-HSIL conductivities.
    Given the simplicity of the model (omitting oligomers and hydroxide anions) and the difficult convergence of theoretical calculations of transport properties like conductivity, the agreement with experiment is remarkable.

    The calculated conductivities in the simulated HSILs result from an average displacement of the net charge resulting from all charge carriers in the system, i.e.  Na$^+$, K$^+$, Rb$^+$ or Cs$^+$ on one hand and Si(OH)$_3$O$^-$ on the other.
    When the movement of ions with opposite charge is correlated, e.g. through the formation of ion pairs, the net transfer of charge reduces, which is reflected in a lower conductivity.
    The Nernst--Einstein approximation of the conductivity completely ignores such correlation as it estimates the conductivity as an average of the diffusion coefficients of the present cations and anions multiplied by their respective charge.\cite{Kubisiak2020}
    Therefore, in reverse reasoning, it can be used to assess the amount of correlated movement in a system by comparing the Nernst--Einstein approximate to the actual conductivity of the simulated systems.

    Comparing Figures~\ref{fig:conductivity} (b) and (c), it is seen that the Nernst--Einstein approximation strongly overestimates the conductivity in all cases, suggesting that cationic and anionic movement are indeed correlated to some extent.
    Importantly, the Nernst--Einstein estimate displays maxima (just like the actual conductivity), namely 10.00 S m$^{-1}$ , 20.61 S m$^{-1}$ , 21.92 S m$^{-1}$  and 21.54 S m$^{-1}$  for Na-HSIL, K-HSIL, Rb-HSIL and Cs-HSIL, respectively.
    The occurrence of these maxima indicates that decreasing ionic conductivity with increasing cationic molar fraction in the ionic liquid regime is not solely attributable to correlated ionic motion.
    If correlated motion were solely responsible for the decrease in conductivity in the ionic liquid regime, the Nernst--Einstein approximation would increase linearly with cation concentration, which is only observed at lower concentrations.

    To further quantify the correlation between cationic and anionic movement, an estimate of the correlated motion is shown in Figure~\ref{fig:conductivity}(d), determined as
    \begin{align}
        X_{\text{corr}} = 1 - \frac{\sigma}{\sigma_{\text{NE}}}.
    \end{align}
    This can be interpreted as the fraction of the estimated Nernst--Einstein conductivity that is missing due to correlated motion of the ions.~\cite{Kubisiak2020}
    It can be seen that the amount of correlation depends strongly on the cation, along the previously discussed series of decreasing charge density.
    While Na-HSIL shows up to 64\% of correlation at its peak, this decreases to 44\%, 35\% and 29\% for K-HSIL, Rb-HSIL and Cs-HSIL, respectively.
    Again, a significant difference is seen between the former and the latter three.
    In any case, the amount of correlation never reaches 1 (as would be expected for a perfectly ion-paired solution) and even decreases at very high concentrations.
    It is therefore clear that correlated motion cannot be the sole explanation for the decreasing conductivity with increasing cationic size.

    \begin{figure}
        \begin{center}
            \includegraphics{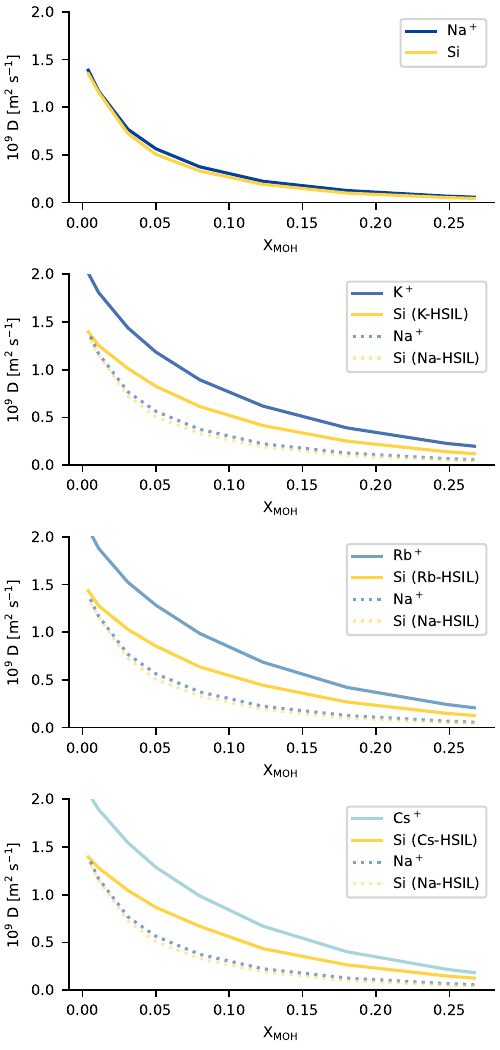}
        \end{center}
        \caption{Diffusion coefficients for the cation (full line) and Si atoms (representing the anion, dashed line) for (a) Na-HSIL, (b) K-HSIL, (c) Rb-HSIL and (d) Cs-HSIL.
        In the latter three, the diffusion coefficient for sodium and silicon (in Na-HSIL) is replotted in a dimmed, dashed line for reference. Error bars are one order of magnitude smaller than data points and thus omitted.}
        \label{fig:diffusion}
    \end{figure}

    An alternative explanation is offered in Figure~\ref{fig:diffusion}, where the diffusion coefficients are given for the cations and anions in the different considered HSIL systems.
    Note the excellent agreement between theoretical results at a cationic mole fraction of 0.004 and experimental results in water at infinite dilution, i.e. 1.39 m$^2$ s$^{-1}$ vs. 1.334 m$^2$ s$^{-1}$ for Na$^+$, 2.01 m$^2$ s$^{-1}$ vs. 1.957 m$^2$ s$^{-1}$ for K$^+$, 2.09 m$^2$ s$^{-1}$ vs. 2.072 m$^2$ s$^{-1}$ for Rb$^+$ and 2.07 m$^2$ s$^{-1}$ vs. 2.056 m$^2$ s$^{-1}$ for Cs$^+$.\cite{Haynes2016}
    For all systems, it is seen that the diffusion coefficient for both the cation and anion drops significantly with increasing cationic concentration as was previously observed for concentrated NaOH solutions.~\cite{Shao2020a,Hellstrom2018}
    More specifically, the diffusion coefficient of the cations at a cationic mole fraction of 0.267 drops to 0.06 m$^2$ s$^{-1}$ for Na-HSIL, to 0.20 m$^2$ s$^{-1}$ for K-HSIL, to 0.21 m$^2$ s$^{-1}$ for Rb-HSIL and to 0.18 m$^2$ s$^{-1}$ for Cs-HSIL.
    This means that movement is much more restricted at higher concentrations, suggesting that mainly the reduced mobility of the individual charge carriers inhibits the conductivity at high concentration, as opposed to correlated motion at low concentration.

    Clear differences are seen between the self-diffusion constants in the Na-HSIL system on one hand and the K-HSIL, Rb-HSIL and Cs-HSIL systems on the other.
    First, the mobility of the charge carriers is much lower in the former system than in the latter three.
    For the aqueous system, it has been reported in the literature that sodium moves slower through water than larger cations, such as cesium, due to its larger hydration shell.\cite{Bourg2010}
    However, it seems that the high charge density also restricts rapid movement in the ionic liquid, compared to cations with lower charge density.
    Second, the cations and anions have very similar diffusion coefficients in Na-HSIL, while the cations are clearly more mobile than the anions in the remaining three systems.
    This reflects the lower extent of correlated movements observed in the case of K-HSIL, Rb-HSIL and Cs-HSIL.

    \begin{figure}
        \begin{center}
            \includegraphics{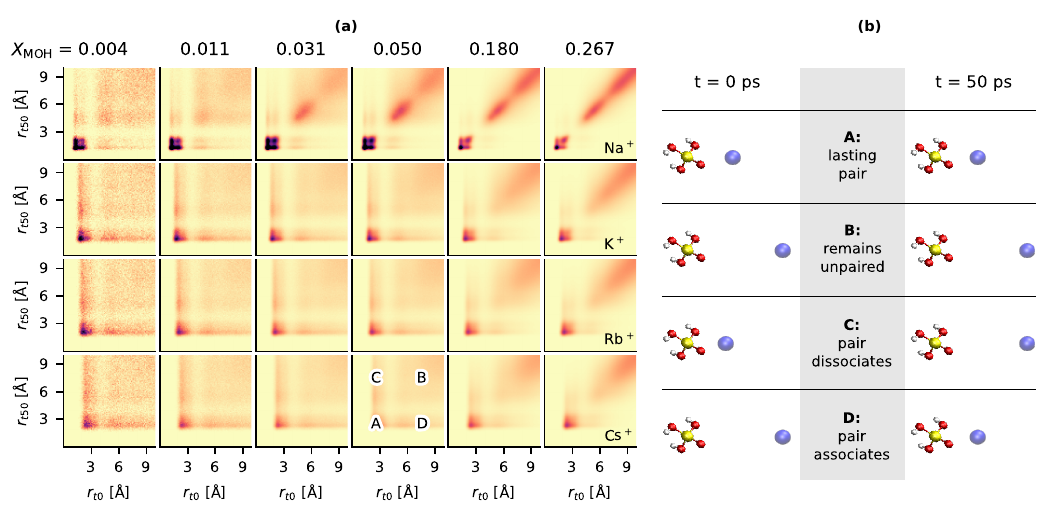}
        \end{center}
        \caption{
            (a) 2D-histograms for Na-HSIL, K-HSIL, Rb-HSIL and Cs-HSIL showing the correlation between a cation-anion distance at a time $t_0$ (x-axis) and the distance for the same pair at a time $t_0 + 50\,\text{ps}$ (y-axis).
            The plots are normalized towards the total number of possible ion pairs for each individual simulation.
            (b) A pictorial clarification of the ion displacements in the respective points indicated in (a) at 0 ps and 50 ps, blue = cesium, yellow = silicon and red = oxygen, red = oxygen and white = hydrogen.
        }
        \label{fig:pairs}
    \end{figure}

    The above discussion clearly demonstrates that conductivity is influenced by the correlated motion of anions and cations, but that this correlation cannot be the sole reason for the reduced conductivity in the limit of high concentrations.
    To understand the nature of the correlated motion, cation-anion distances were calculated at time frames separated by 50 ps to assess the prevalence of lasting ion pairs.
    Figure~\ref{fig:pairs}a shows normalized (towards the total number of possible ion pairs for the given simulation) 2D-histograms of these distances at snapshots separated by $\Delta t$ = 50 ps for all HSIL systems at several concentrations.
    The figure thus serves as a heatmap representing the occurrence of a certain interionic distance at $t_0$ + 50 ps, given a specific distance at $t_0$.
    In other words, the plots give a measure of the extent to which ions have moved relative to each other within this time frame as shown schematically in Figure \ref{fig:pairs}b.
    For each histogram in (a), the lower left corner (point A) corresponds to directly interacting pairs that stay in place (or break and form again) over 50 ps, while a point on the diagonal (point B) corresponds to a longer distance that has not changed over 50 ps.
    A point above (below) the diagonal, like point C (point D), indicates a distance that has increased (decreased) over 50 ps.
    Hence, a high probability density along the diagonal indicates little change in cation-anion distance over a 50 ps time interval, while a significant off-diagonal density corresponds to large relative motion of opposite ions.
    Several lag times were considered, whereby 50 ps was found to be most illustrative for the differences in ion pair formation between Na-HSIL on one hand and K-HSIL, Rb-HSIL and Cs-HSIL on the other, as clearly observed in Figure~\ref{fig:pairs}.

    For Na-HSIL, lasting (at least 50 ps) ion pairs are clearly observed in the aqueous regime as evidenced by the occurrence of a strongly highlighted area at short distances.
    As the concentration increases, the amount of ion pairs increases along the aqueous regime, yet decreases along the ionic liquid regime.
    This decrease does not imply that no ion aggregation is occurring at higher concentration (in fact, ion aggregation is inevitable), but it does mean that their interactions are much shorter lived (less than 50 ps) than in the aqueous regime.
    This will be further elaborated upon in the next section (section~\ref{sec:liquid_structure}) where the liquid structure is discussed in detail.
    Interestingly, in the ionic liquid regime unpaired cation-anion combinations are observed along the diagonal only, meaning that their interatomic distance changes little over 50 ps even though they are not in close contact.
    This is related to the reduced mobility of the ions (lower diffusion coefficients, see above), whereby limited movement is observed within the investigated time frame.

    For the remaining three cations at low concentrations, the ion-paired area is less densely populated and counts are almost equally distributed in space.
    This confirms the previous results as decreased charge density leads to less lasting ion pairs and thus decreased correlated motion.
    As the concentration increases, the observed distribution becomes narrower centered around the diagonal, confirming the observed decreasing diffusion coefficients.
    However, as the diffusion coefficients remain higher than for Na-HSIL, the distribution around the diagonal is also broader with increasing cationic radius.
    It is further interesting to note that in all simulated systems, the transition from an aqueous regime to an ionic liquid regime is very continuous and is not associated with an abrupt change in ion pair formation.

    It is interesting to highlight the systematic trends observed between the discussed results and the ionic radii of the cations, in line with the results of Mahler et al.\cite{Mahler2012}
    It is seen that with decreasing charge density (increasing cationic radius), the conductivity and mobility increase, while the amount of correlated motion decreases.
    This suggests that the conductivity is an interplay between correlated motion and charge carrier mobility, whereby the relative importance of both effects depends on the charge density of the cation.
    It seems that larger charge density leads to strongly correlated motions, yet slower diffusion, while for increasing cation size, the correlated motion drops and diffusion coefficients rise.
    Finally, it is important to emphasize the qualitative nature of the observations made as a result of the simplicity of the used model.
    While the assumption of singly deprotonated monomers is validated by the reasonable agreement with experimental results, it is unavoidable that this assumption has an effect on the simulated ion aggregation.
    In fact, experimental results have shown that silicates are doubly protonated at Si/OH ratios lower than 1 (validating the model in current work), yet the degree of polymerization increases with increasing Si/OH ratio.\cite{Asselman2023}
    More complex models are currently being developed to account for this in future work.

    \subsection{Liquid Structure}
    \label{sec:liquid_structure}

    \begin{figure}
        \begin{center}
            \includegraphics{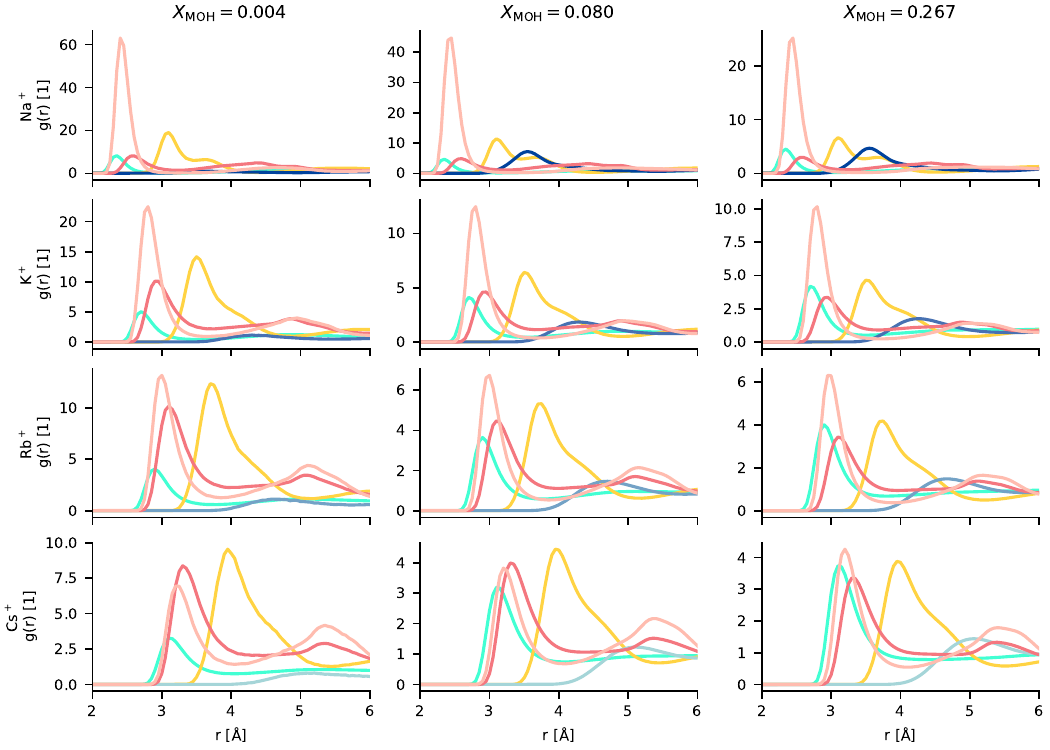}
        \end{center}
        \caption{Radial distribution functions for Na-HSIL, K-HSIL, Rb-HSIL and Cs-HSIL, from top to bottom respectively, at cationic mole fractions of 0.004 (left), 0.080 (middle) and 0.267 (right).
        4 shades of blue are used for the respective cations, light blue for the water oxygen, yellow for silicon, light red for the deprotonated oxygen and dark red for the protonated oxygens.}
        \label{fig:RDFs}
    \end{figure}

    Radial distribution functions were calculated between selected species of interest to gain an understanding of the liquid structure of HSILs as a function of cation type and concentration.
    More specifically, the oxygen atom of water, the cations, the silicon atom and its oxygens were taken into account.
    Thereby, a distinction was made between protonated and deprotonated oxygen atoms on the asymmetric silicate anion.
    Figure~\ref{fig:RDFs} shows the radial distribution functions around the cation in the different considered HSIL systems at three different cationic mole fractions.
    More specifically, the radial distribution functions are shown between the respective cations and the remaining species in the simulation, namely water, silicon, the cation itself and the protonated and deprotonated oxygens in the silicate monomers.
    The shown concentrations correspond to the lowest (aqueous regime), intermediate and highest (ionic liquid regime) cationic mole fractions considered, i.e. $X_\text{NaOH}$ = 0.004, 0.080 and 0.267.
    Notably, the intermediate mole fraction corresponds to the transition region between aqueous and ionic liquid regimes.

    Starting with Na-HSIL at the lowest cationic concentration, a low and relatively broad peak is found for water with a maximum at 2.35 \AA.
    This is closely followed by a high peak at 2.40 \AA\ representing the deprotonated oxygen of the silicate anion and a low and broad peak at 2.60 \AA\ for the protonated oxygens on the silicate monomer.
    Note that there are three protonated and one deprotonated oxygen for each monomer.
    The large peak for the deprotonated oxygen confirms the preference of the negatively charged oxygen and the Na$^+$ cation to form ion pairs.
    Finally, there is an interesting peak for the silicon atom, which displays a maximum at 3.10 \AA\ and a shoulder at 3.65 \AA.
    This indicates two different orientations of the silicate monomer relative to the cation, suggesting that the silicate anion can be oriented with either the deprotonated or protonated oxygen, respectively, towards the cation.
    The strong and sharp peak indicates that the former orientation is preferred and very directional, which is intuitive given the positive and negative charges involved.
    The lower, broader peaks of the Si shoulder and protonated oxygen indicate that this orientation is less pronounced.

    With increasing cation concentration, this general pattern is maintained and the maxima of the peak remain at the same location.
    However, the peaks associated with the silicate monomer (silicon and both oxygen types) get weaker with increasing ionic concentrations.
    This may seem counterintuitive and in contradiction to the results presented above at first, but it should be noted that radial distribution functions are a ratio of the local distribution of neighbors over the uniform distribution.
    Because the (average) concentration appears in the denominator, any tendency to local attraction at low cationic concentration will result in high large peaks.
    In contrast, at high concentrations, the difference between the local distribution and the uniform average is reduced.
    This will become clearer when considering coordination numbers below.
    Importantly, in the ionic liquid regime at higher cation concentrations, a peak is observed at 3.55 \AA\ for sodium that is not observed in the aqueous regime.
    This is a tell-tale sign of an ionic liquid, whereby the positive cations show long-range ordering within the liquid.\cite{Salanne2015}
    This coincides with experimental results showing that the ionic liquid regime is entered around the maximum measured conductivity.\cite{Pellens2022}
    Yet, it should be noted that the liquid is not overly structured, i.e. only a second cation---cation peak is observed and not a third.

    When comparing the remaining cations with sodium, it is immediately clear that the peaks shift to larger distances with increasing cationic radii.
    More importantly, the intensity of the peak with the deprotonated oxygen becomes much less pronounced (relative to the Si and protonated oxygen peaks) with increasing cation size.
    As a result, also the shoulder of the radial distribution function with Si becomes much less pronounced, while for Cs the deprotonated and protonated oxygen have peaks of similar heights.
    This confirms that with increasing cation size, thus decreasing charge density, the tendency to form ion pairs diminishes and the silicate monomer retains much more flexibility to rotate within formed ion pairs.
    Together with the less pronounced cation---cation peaks, this shows that the K-HSIL, Rb-HSIL and Cs-HSIL liquids are increasingly less structured than Na-HSIL, following the charge density trend.
    Finally, it is interesting to highlight the correspondence between the calculated radial distribution functions and experimental measurements on MOH systems.
    More specifically, the maxima of the calculated radial distribution funcion for Na--O$_{\text{water}}$ of 2.35 \AA\ compares very well with the experimental value of 2.43 \AA, for K--O$_{\text{water}}$, 2.80 vs. 2.81 \AA, for Rb--O$_{\text{water}}$ 2.90 \AA\ vs. 2.98 \AA\ and for Cs--O$_{\text{water}}$ 3.1 \AA\ vs. 3.07 \AA.\cite{Mahler2012}

    \begin{figure}
        \begin{center}
            \includegraphics{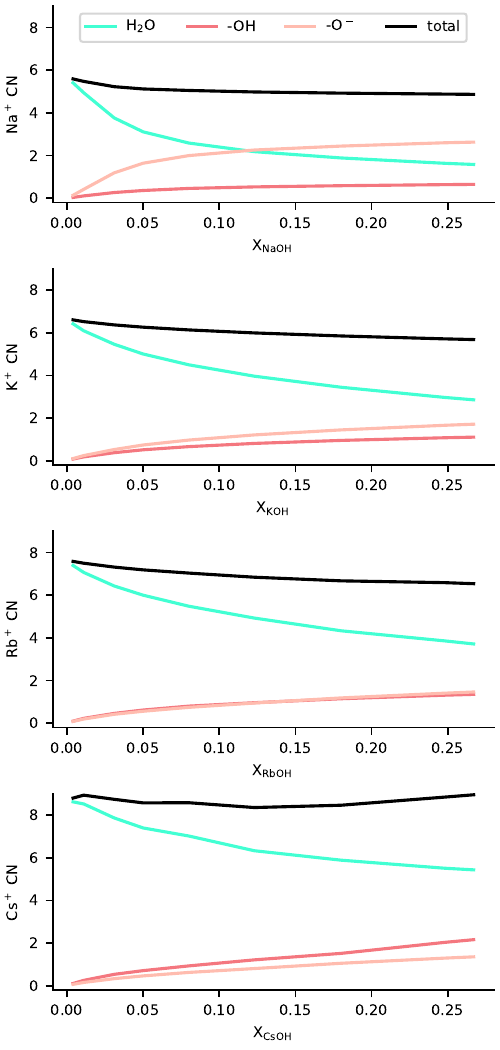}
        \end{center}
        \caption{Coordination numbers for water, cation and the deprotonated and protonated oxygens as well as the total coordination around the alkali cation within the Na-HSIL, K-HSIL, Rb-HSIL and Cs-HSIL systems, from top to bottom respectively.}
        \label{fig:coord_num}
    \end{figure}

    From experimental results, it was suggested that, in the ionic liquid regime, water stops acting as a solvent and rather acts as a ligand to the cation due to the low water content.\cite{Pellens2022}
    To assess this hypothesis, the presented radial distribution functions were integrated to obtain the absolute number of molecules of water, protonated oxygen and deprotonated oxygen present in the respective coordination spheres of the studied cations as represented in Figure~\ref{fig:coord_num}.
    Given the proximity of the maxima of these coordination numbers, the corresponding molecules may be considered as the first solvation shell of the respective cations, which is also plotted in the figure.
    However, as ``solvation shell'' is a poor description in the context of ionic liquids, the term ``first coordination shell'' will be used instead.

    From Figure~\ref{fig:coord_num}, it can be seen that the cations are entirely solvated by water at lowest ionic concentration leading to hydration numbers in close correspondence with experiments of alkali ions in aqueous solution, namely 5.4 (exp: 5.7), 6.4 (6.9), 7.4 (8.0) and 8.6 (8.0) for Na$^+$, K$^+$, Rb$^+$ and Cs$^+$, respectively.\cite{Mahler2012}
    Note that this is not in contradiction with the long-lasting ion pairs observed in Figure~\ref{fig:pairs} as these ion pairs are fully solvated, leading to a slightly reduced hydration number only.
    As the cationic concentration increases, the first coordination sphere starts losing water, which is gradually replaced by oxygens from the silicate anions.
    The number of species (water, protonated and deprotonated oxygen atoms) in the first coordination shell remains remarkably constant through the entire concentration range.
    This supports the understanding that the cations are competing for water molecules and complete their coordination spheres with silicate anions due to a lack of water molecules.\cite{Asselman2021}
    Interestingly, a clear difference is once again observed between Na$^+$ on one hand and K$^+$, Rb$^+$ and Cs$^+$ on the other.
    Thereby, the former shows a clear preference to replace water molecules by deprotonated oxygens, while cesium at the other side of the charge density scale, takes up more protonated than deprotonated oxygens in its first coordination shell.
    As an example, the coordination numbers for the deprotonated oxygen and protonated oxygens in the systems with highest cationic mole fraction are 2.63 and 0.65 for sodium, respectively, while for K, Rb and Cs these coordination numbers are 1.72 and 1.1, 1.47 and 1.36 and finally 1.36 and 2.16.
    This is in line with the radial distribution functions as well as the observed amount of correlated motion described above.
    Note that the ``preference'' of Cs$^+$ for protonated oxygen should be interpreted in the knowledge that there are three times more protonated than deprotonated oxygens.

    \begin{figure}
        \begin{center}
            \includegraphics{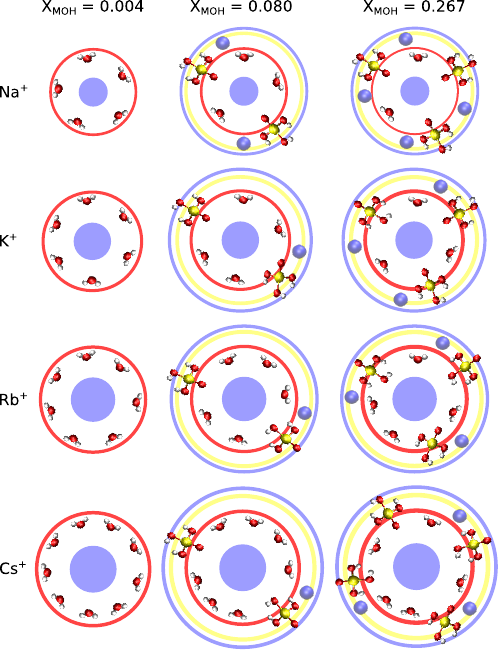}
        \end{center}
        \caption{Intuitive 2D-models of the first solvation shell of the alkali cation in the Na-HSIL, K-HSIL, Rb-HSIL and Cs-HSIL systems, from top to bottom respectively, at cationic mole fractions of 0.004 (left), 0.080 (middle) and 0.267 (right).
        The circles indicate the first minimum of the respective radial distribution functions for oxygen (red), silicon (yellow) and the alkali cation (blue).
        The central cation and the radial distribution function minima are at scale, the other atoms are not in order to avoid cluttering.}
        \label{fig:2D}
    \end{figure}

    From the coordination numbers and the radial distribution functions, an intuitive 2D-model can be derived of the structure of the liquid as shown in Figure~\ref{fig:2D} for the same concentrations discussed previously.
    Thereby, the concentric circles represent the first minima of the respective radial distribution functions or in other words the sphere within which the respective atoms are located on average.
    The coordination numbers of the separate species suggest the appropriate number (which must be rounder for this exercise) to be positioned within these circles.
    Interestingly, the two orientations of the silicate anion can be understood and quantified through the coordination numbers of the protonated and deprotonated oxygens oriented towards the cation.
    It is now very clear that sodium at a mole fraction of 0.267 has three silicate anions in its first coordination sphere, three of which point their deprotonated oxygens towards the cation.
    Note that this is an average number over the simulation time and that in reality the silicates will be rotating to break and reform ionic interactions as to maintain charge neutrality locally.
    For Cs on the other hand, four silicate anions are surrounding the cation, only one of which is directly pointing a deprotonated oxygen towards it.
    Given that there are three different protonated oxygens, this suggests that the ionic liquid becomes less structured with decreasing charge density.

    Together with the results presented above, this paints a clear picture of ion pair formation in the two different regimes.
    In the aqueous regime, ion pair formation is observed in the traditional sense of the concept, i.e. formation of exclusive pairs of a single cation and single anion surrounded by solvent that migrate through the liquid together.~\cite{Marcus2006}
    As the amount of charge carriers is increased, the number of ions pairs increases, yet the conductivity does not increase linearly due to the concerted motion of the ion pairs.
    Upon entering the ionic liquid regime, so many ion pairs are formed that further ion aggregation occurs whereby the ion pairs are no longer exclusive, have a much shorter lifetime and display reduced diffusive motion.
    This suggests the formation of a glassy network whereby every ion forms multiple pairwise ionic interactions with surrounding ions of opposite charge, which constantly break and reform.
    The number and exclusiveness of these ionic interactions in both regimes then depend on the charge density of the involved cation.

    Apart from these observations, it is clearly seen for all HSIL systems that the cations bring protonated and deprotonated oxygens in close contact with each other.
    Furthermore, it is easy to imagine spontaneous condensation reactions happening between a protonated and deprotonated oxygen, promoted by the present cations.
    This suggests that the cations play a crucial role in the very first condensation reactions within this mixture, reminiscent to the concept of cation-mediated assembly as described by Cundy and Cox.~\cite{Cundy2005}
    This is important as the influence of cations on the condensation reaction is empirically known, but not understood.\cite{Trinh2009}
    Interestingly, some authors noted that fluoride-containing highly concentrated gels are ``reminiscent of ionic liquids and force close contact between the [organic] cation and silica at the initial stages of the synthesis [\dots] which may be the reason for very successful zeolite formation''.\cite{Rimer2006}

    \section{Conclusions}
    \label{sec:conclusions}

    This work presents a detailed molecular model of hydrated silicate ionic liquids (HSILs), a recently proposed growth liquid for zeolites that removes many of the experimental hurdles to understand zeolite nucleation and growth.
    The conductivity and liquid structure of four different HSILs, Na-HSIL, K-HSIL, Rb-HSIL and Cs-HSIL, is assessed as a function of type of alkali cation and its concentration.
    The different systems are characterized using large scale molecular dynamics simulations on a very simple system including silicate monomers, water and alkali cations.
    Experimental conductivity measurements are reported that show remarkable qualitative agreement with the theoretical results.
    It should be noted that Rb-HSIL was not measured experimentally due to its high cost.
    However, the theoretical predictions are well in line with observed trends for the remaining systems, providing reasonable confidence in the Rb-HSIL models.
    Despite the simplicity of the proposed models, the agreement with experimental results lead to important conclusions on the ion pairing in HSILs, which are important for future investigation, both theoretical and experimental.
    Importantly, the development of realistic molecular models of traditional zeolite synthesis liquids is prohibitively difficult due to their heterogeneous nature.
    As such, this work represents an important step towards a full atomistic description of a realistic growth liquid for zeolite nucleation.

    It is found that the observed conductivity and liquid structure are strongly dependent on the cation's charge density on one hand and on the concentration on the other.
    At lower concentrations, the HSILs behave as an aqueous liquid characterized by increasing conductivity with increasing concentration.
    At a certain point, a maximum conductivity is reached, after which the conductivity drops, identified as the ionic liquid regime in line with prior experimental investigation.
    This work shows that the decreasing conductivity is governed by an interplay between increasing ion pair formation and decreasing diffusion coefficients.
    The extent to which both factors contribute is strongly dependent on the type of cation as the decreasing charge density along the Na, K, Rb and Cs series, leads to a decreasing tendency to form ion pairs as well as slower diffusion.
    Furthermore, the ionic liquid becomes less structured at lower charge density.
    Despite this, it is found that for all HSILs, the cation brings multiple silicate monomers in close contact, possibly promoting initial oligomerisation reactions between silicate monomers.
    Further research using more complicated and reactive molecular models is required to confirm this hypothesis, work in that direction is currently performed.
    The slow diffusion and the pairing of each ion with multiple counterions at higher concentrations, suggest that zeolite formation occurs in a glassy network within this HSIL regime.

    \section*{Supporting Information}
    A PDF document containing a table with the exact compositions of all performed simulations as well as a table containing their cell vectors.
    A ZIP file with the force field XML file that is directly usable in OpenMM.

    \begin{acknowledgement}
    J.V. acknowledges the Research Foundation—Flanders (FWO) for a junior postdoctoral mandate (Project No.\ 12E6423N).
    T.V. thanks the Research Board of Ghent University (BOF) for its financial support.
    This research was supported by a joint Research Foundation-Flanders (FWO) (Project No.\ G083318N) and Austrian Science Fund (FWF) (project ZeoDirect I3680-N34) grant.
    The computational resources and services used in this work were provided by the VSC (Flemish Supercomputer Center), funded by the Research Foundation—Flanders (FWO) and the Flemish Government.
    NMRCoRe is supported by the Hercules Foundation (AKUL/13/21), by the Flemish Government as an international research infrastructure (I001321N) and by department EWI via the Hermes fund (AH.2016.134).
    \end{acknowledgement}

    \bibliography{references}

\end{document}


\begin{table}
        \begin{center}
        \begin{tabular}{l|c c c c}
            X$_\text{MOH}$ & $n$ & H$_2$O & M$^+$ & Si(OH)$_3$O$^-$ \\
            \hline
            0.004 & 260.5 & 10458 & 40 & 40 \\
            0.011 & 90 & 10918 & 120 & 120 \\
            0.031 & 31 & 10316 & 320 & 320 \\
            0.050 & 18.5 & 9351 & 480 & 480 \\
            0.080 & 11 & 8635 & 720 & 720 \\
            0.123 & 6.6 & 8206 & 1080 & 1080 \\
            0.180 & 4 & 7224 & 1440 & 1440 \\
            0.248 & 2.5 & 6442 & 1840 & 1840 \\
            0.267 & 2.2 & 6181 & 1920 & 1920 \\
        \end{tabular}
        \end{center}
        \caption{Number of water molecules, $n$, present in the simulations with molar compositions of H$_2$O:SiO$_2$:MOH $n$:1:1 as well as the actual number of water molecules (H$_2$O), metal cations (M$^+$) and silicate monomers (Si(OH)$_3$O$^-$) in the simulation box}
        \label{tab:compositions}
    \end{table}

    \begin{table}
        \begin{center}
        \begin{tabular}{l|c c c c}
            X$_\text{MOH}$ & Na-HSIL & K-HSIL & Rb-HSIL & Cs-HSIL \\
            \hline
            0.004 & 56.744 & 56.785 & 56.727 & 56.748  \\
            0.011 & 58.441 & 58.589 & 58.804 & 58.747  \\
            0.031 & 59.571 & 60.096 & 60.276 & 60.433  \\
            0.050 & 59.933 & 60.404 & 60.804 & 61.124  \\
            0.080 & 61.257 & 61.904 & 62.649 & 62.959  \\
            0.123 & 64.178 & 65.268 & 66.056 & 66.482  \\
            0.180 & 66.113 & 67.454 & 68.491 & 69.048 \\
            0.248 & 68.424 & 70.110 & 71.317 & 72.084 \\
            0.267 & 68.730 & 70.394 & 71.719 & 72.486 \\
        \end{tabular}
        \end{center}
        \caption{Cell vectors in nm of the simulation boxes used in this work.}
        \label{tab:cell vectors}
    \end{table}